\documentclass[twocolumn,nofootinbib,superscriptaddress,preprintnumbers,floatfix]{revtex4}

\usepackage{graphicx}
\usepackage{amsmath}
\usepackage[bookmarks=false]{hyperref}

\newcommand{\eq}[1]{Eq.~\eqref{eq:#1}}
\newcommand{\eqs}[2]{Eqs.~\eqref{eq:#1} and \eqref{eq:#2}}
\renewcommand{\sec}[1]{Sec.~\ref{sec:#1}}
\newcommand{\subsec}[1]{Sec.~\ref{subsec:#1}}
\newcommand{\fig}[1]{Fig.~\ref{fig:#1}}

\DeclareMathOperator{\Repart}{Re}

\renewcommand{\Re}{\Repart}

\newcommand{\abs}[1]{\lvert#1\rvert}
\newcommand{\ABs}[1]{\Bigl\lvert#1\Bigr\rvert}
\newcommand{\ABS}[1]{\biggl\lvert#1\biggr\rvert}
\newcommand{\mae}[3]{\langle#1\lvert#2\rvert#3\rangle}
\newcommand{\vevB}[1]{\langle #1 \rangle_{\! B}}
\newcommand{\ket}[1]{\lvert#1\rangle}
\newcommand{\ord}[1]{\mathcal{O}(#1)}

\newcommand{\df}{ \mathrm{d} }
\newcommand{\img}{\mathrm{i}}

\newcommand{\w}{\omega}
\newcommand{\e}{\varepsilon}

\newcommand{\cH}{\mathcal{H}}

\newcommand{\hC}{\widehat{C}}
\newcommand{\hP}{\widehat{P}}
\newcommand{\hF}{\widehat{F}}

\newcommand{\bn}{{\bar n}}
\newcommand{\vslash}{v\!\!\!\slash}
\newcommand{\nslash}{n\!\!\!\slash}
\newcommand{\bnslash}{\bar n\!\!\!\slash}

\newcommand{\GeV}{ {\:\mathrm{GeV}} }

\newcommand{\nn}{\nonumber}

\newcommand{\lqcd}{\Lambda_\mathrm{QCD}}
\newcommand{\cut}{\mathrm{cut}}
\newcommand{\incl}{\mathrm{incl}}
\newcommand{\mix}{\mathrm{mix}}

\newcommand{\pxp}{\ensuremath{p_X^+}}

\newcommand{\ellpm}{ \ell^+ \ell^- }
\newcommand{\mbbar}{\overline m_b}

\newcommand{\C}{ \mathcal{C} }

\def\babar{\mbox{\sl B\hspace{-0.4em} {\small\sl A}\hspace{-0.37em} \sl
B\hspace{-0.4em} {\small\sl A\hspace{-0.02em}R}\hspace{0.3em}}}

\allowdisplaybreaks[4]

\begin{document}


\preprint{\vbox{\hbox{arXiv:0812.0001}\hbox{CALT-68-2710}\hbox{MIT--CTP 3999}}}

\title{\boldmath Nonperturbative $m_X$ cut effects in $B\to X_s\,\ell^+\ell^-$ observables}

\author{Keith S.\ M.\ Lee}
\affiliation{California Institute of Technology, Pasadena, CA 91125}

\author{Frank J.\ Tackmann}
\affiliation{Center for Theoretical Physics, Massachusetts Institute of
Technology, Cambridge, MA 02139}

\begin{abstract}

Recently, it was shown that in inclusive $B\to X_s\,\ell^+\ell^-$ decay, an
angular decomposition provides three independent ($q^2$ dependent)
observables. A strategy was formulated to extract all measurable
Wilson coefficients in $B\to X_s\,\ell^+\ell^-$ from a few
simple integrals of these observables in the low $q^2$ region. The
experimental measurements in the low $q^2$ region require a cut on the
hadronic invariant mass, which introduces a dependence on nonperturbative
$b$ quark distribution functions. The associated hadronic uncertainties
could potentially limit the sensitivity of these decays to new physics.
We compute the nonperturbative corrections to all three observables at leading
and subleading order in the power expansion in $\lqcd/m_b$. We find that the
subleading power corrections give sizeable corrections, of order $-5\%$ to
$-10\%$ depending on the observable and the precise value of the hadronic
mass cut. They cause a shift of order $-0.05\GeV^2$ to $-0.1\GeV^2$
in the zero of the forward-backward asymmetry.

\end{abstract}

\maketitle

\section{Introduction}
\label{sec:intro}

The inclusive decay $B\to X_s\,\ell^+\ell^-$ is highly sensitive
to new physics, since it involves flavor-changing neutral-current interactions,
which do not occur at tree level in the standard model (SM). It is
described by the effective Hamiltonian
\begin{equation}\label{eq:Heff}
\cH_\mathrm{eff} = -\frac{G_F}{\sqrt2}\, V_{tb} V^*_{ts}\,\sum_{i=1}^{10} C_i\, O_i
\,,\end{equation}
where $O_{1-6}$ are four-quark operators and
\begin{align}
O_7 &= \frac{e}{4\pi^2}\,\mbbar
   \, \bar s\, \sigma_{\mu\nu}F^{\mu\nu} P_R\, b
\,,\nn\\
O_8 &= \frac{g}{4\pi^2}\, \mbbar
  \, \bar s\,\sigma_{\mu\nu} G^{\mu\nu} P_R\, b
\,,\nn\\
O_9 &= \frac{e^2}{4\pi^2}\,
  (\bar s\, \gamma_\mu P_L b)\, (\bar\ell \gamma^\mu \ell)
\,,\nn\\
O_{10} &= \frac{e^2}{4\pi^2}\,
  (\bar s\, \gamma_\mu P_L b)\, (\bar\ell \gamma^\mu\gamma_5 \ell)
\,,\end{align}
with $P_{L,R} = (1\mp\gamma_5)/2$. Here we have neglected the $s$-quark mass.
All short distance information, including possible new physics contributions,
is encoded in the Wilson coefficients, $C_i$. Thus, one can test the SM and
search for new physics by extracting these Wilson coefficients. Two
observables frequently studied for this task are the dilepton mass ($q^2$)
spectrum and the forward-backward asymmetry~\cite{Grinstein:1988me, Ali:1991is, Cho:1996we}.
Recently, it was noted that a third observable, proportional to a different
combination of Wilson coefficients, can be obtained from an angular
decomposition and significantly increases the sensitivity
to the different Wilson coefficients~\cite{Lee:2006gs}. With the addition of this
third observable, the precise measurement of the $q^2$ dependence becomes
unnecessary. Instead, a few simple $q^2$ integrals of these observables are
sufficient to determine all measurable Wilson coefficients in
$B\to X_s\,\ell^+\ell^-$ with the data already available from the
$B$ factories.

The presence of intermediate $c\bar c$ resonances, $J/\psi$ and $\psi'$,
restricts the portion of phase space that is amenable to a precise
comparison between theory and experiment. There are two suitable
regions, $q^2 < m_{J/\psi}^2$ and $q^2 > m_{\psi'}$. The large $q^2$
region is usually considered less favorable, because it has a smaller rate
and suffers from large nonperturbative corrections. However, the experimental
efficiency is better there, and in Ref.~\cite{Ligeti:2007sn} it was
shown that by taking the ratio of the $B\to X_s\,\ell^+\ell^-$ and
$B\to X_u\ell\bar\nu$ rates the nonperturbative corrections can be kept under
control, so precise predictions are possible even at large $q^2$.

In this paper, we focus on the low $q^2$ region, which benefits from a
higher rate. The trade-off is that experimentally a cut on the hadronic
invariant mass, $m_X < m_X^{\rm cut}$, is required to suppress
the huge background from $b\to c(\to s \ell^+ \nu) \ell^-\bar\nu$.
This $m_X$ cut introduces hadronic uncertainties that can easily spoil the
search for new physics in this decay. The problem is that the decay rate
is put into a kinematic region where the usual local operator product
expansion in powers of $\Lambda_{\rm QCD}/m_b$ is no longer applicable.
Instead, the rate becomes sensitive to the motion of the $b$ quark inside
the $B$ meson, which is described by nonperturbative $b$ quark
distribution functions (shape functions)~\cite{Neubert:1993ch, Bigi:1993ex}.
(The large $q^2$ region is unaffected by the $m_X$ cut.)

The latest \babar~\cite{Aubert:2004it} and Belle~\cite{Iwasaki:2005sy}
analyses use $m_X^\cut = 1.8\GeV$ and $m_X^\cut = 2.0\GeV$, respectively.
To obtain the total branching ratio in the low $q^2$ region, the measurements
are extrapolated to the full $m_X$ range using signal Monte Carlo based on a
Fermi motion model. The extrapolated measurements are routinely quoted
to compare theory and experiment. This practice is questionable because,
as is well established in the context of inclusive $B\to X_s\gamma$ and
$B\to X_u\ell\bar\nu$ decays, any such extrapolation should not be considered
reliable and can give at best a rough estimate of the effect of the $m_X$ cut.

In the kinematic region of low $q^2$ and small $m_X$, one can calculate
the inclusive decay rates using soft-collinear effective theory
(SCET)~\cite{Bauer:2000ew, Bauer:2000yr}. At leading order in $\lqcd/m_b$,
they factorize into process-dependent hard functions $h^{[0]}$, a universal
jet function $J$, and the universal soft shape function
$S$~\cite{Korchemsky:1994jb, Bauer:2001yt}, i.e.\
\begin{equation}\label{eq:fact}
\df\Gamma^{[0]} = h^{[0]}\times J\otimes S
\,,\end{equation}
a result applied extensively in the study of inclusive $B\to X_u\ell\bar\nu$ and
$B\to X_s\gamma$ decays. It was first applied to $B\to X_s\,\ell^+\ell^-$ in
Refs.~\cite{Lee:2005pk, Lee:2005pw} to study systematically the effect of
the $m_X^\cut$ on the $q^2$ spectrum and forward-backward asymmetry. In
Ref.~\cite{Lee:2005pw} it was shown that the cut on $m_X$ leads to a $10-30\%$
reduction in the rate. This reduction is, to a good approximation, universal
among the different short distance contributions and one can take it into
account accurately using experimental information from $B\to X_s\gamma$ or
$B\to X_u\ell\bar\nu$, thereby maintaining the sensitivity to new physics.

The largest irreducible hadronic uncertainties and universality breaking
are expected to come from $\ord{\lqcd/m_b}$ power corrections due to
subleading shape functions~\cite{Bauer:2001mh, Leibovich:2002ys, Bauer:2002yu}.
In this paper, we extend the analysis of the three angular observables to
incorporate nonperturbative shape-function effects arising from the $m_X$ cut,
including the $\ord{\lqcd/m_b}$ subleading shape functions.

In \sec{general}, we briefly discuss the kinematics and the angular
decomposition, defining the three observables $H_{T,A,L}(q^2)$. In
\sec{splitmatching}, we discuss the separation of the perturbation series
above and below the scale $\mu\sim m_b$, and our effective Wilson coefficients.
In \sec{results}, we present our results for $H_{T,A,L}$ in the SCET region.
The leading power contribution is given in \subsec{LO}, including the full
NLL and partial NNLL perturbative corrections. The subleading power
corrections are presented at tree level in \subsec{NLO}. Their numerical
impact is investigated briefly in \sec{SSFeffects}, and we conclude in
\sec{conclusions}.

\section{Angular Decomposition and Kinematics}
\label{sec:general}

The triple differential decay rate can be written as~\cite{Lee:2006gs}
\begin{align}\label{eq:d3Gamma}
\frac{\df^3\Gamma}{\df q^2\, \df\pxp\, \df z}
&= \frac{3}{8} \bigl[(1 + z^2) H_T(q^2,\pxp) +  2 z H_A(q^2,\pxp)
\nn\\
& \qquad + 2(1 - z^2) H_L(q^2,\pxp) \bigr]
\,.\end{align}
Here, $q^2 = (p_{\ell^+} + p_{\ell^-})^2$ is the dilepton invariant mass, $p_X^\pm = E_X \mp \lvert\vec{p}_X\rvert$, and $z = \cos\theta$.
In $\bar B^0$ or $B^-$ [$B^0$ or $B^+$] decay, $\theta$ is the angle between the
$\ell^+$ [$\ell^-$] and the $B$ meson three-momenta in the $\ellpm$ center-of-mass
frame. The $q^2$ spectrum and forward-backward asymmetry are given by
\begin{equation}
\frac{\df\Gamma}{\df q^2} = H_T(q^2) + H_L(q^2)
\,,\qquad
\frac{\df A_\mathrm{FB}}{\df q^2} = \frac{3}{4}\,H_A(q^2)
\,.\end{equation}

The velocity of the $B$ meson is $v^\mu = p_B^\mu/m_B$. We define
light-cone vectors $n$ and $\bar{n}$ such that $q_\perp^\mu = v_\perp^\mu = 0$
and $\pxp = n\cdot p_X$, $p_X^- = \bn\cdot p_X$.
For later convenience, we also define the leptonic light-cone variables
\begin{align}
q_+ &= n\cdot q = m_B - \pxp
\,,\nn\\
q_- &= \bn \cdot q = m_B - p_X^- = \frac{q^2}{m_B - \pxp}
\,,\end{align}
with $q^2 = q_+ q_-$.

The functions $H_i(q^2,\pxp)$ in \eq{d3Gamma} are independent of $z$, and are
given by
\begin{align}\label{eq:HTAL}
H_T(q^2,\pxp) &= 2\frac{\Gamma_0}{m_B^5}\, \frac{(q_+ - q_-)^2}{q_+}\,q^2\, W_T(q^2, \pxp)
\,, \nn \\
H_A(q^2,\pxp) &= -2\frac{\Gamma_0}{m_B^5}\,\frac{(q_+ - q_-)^2}{q_+}\,q^2\, W_A(q^2, \pxp)
\,, \nn \\
H_L(q^2,\pxp) &= \frac{\Gamma_0}{m_B^5}\,\frac{(q_+ - q_-)^2}{q_+}\, W_L(q^2, \pxp)
\,,\end{align}
where
\begin{equation}
\Gamma_0  =  \frac{G_F^2\,m_B^5}{48\pi^3}\,\frac{\alpha_{\rm em}^2}{16\pi^2}\, |V_{tb} V_{ts}^*|^2\,.
\end{equation}
In terms of the usual structure functions in the decomposition of the hadronic tensor,
\begin{align}\label{eq:Wmunu}
W^{\mu\nu} &= \frac{1}{2m_B}\,\frac{1}{2\pi} \int\!\df^4x\, e^{-\img q\cdot x} \mae{B}{J^{\dagger\mu}(x)\, J^\nu(0)}{B}
\nn\\
&=
- g^{\mu\nu} W_1 + v^\mu v^\nu W_2 + \img\epsilon^{\mu\nu}{}_{\alpha\beta} v^\alpha q^\beta W_3
\nn\\ & \quad
+ q^\mu q^\nu W_4 + (v^\mu q^\nu + v^\nu q^\mu) W_5
\,,\end{align}
the hadronic structure functions $W_{T,A,L}$ in \eq{HTAL} are given by
\begin{align}\label{eq:WTAL}
W_T &= 4\, W_1
\,, \nn \\
W_A &= -2\,(q_+ - q_-)\, W_3
\,, \nn \\
W_L &= 4\,q^2\,W_1 + (q_+ - q_-)^2\, W_2
\,.\end{align}

\begin{figure}[t]
\includegraphics[width=0.66\columnwidth]{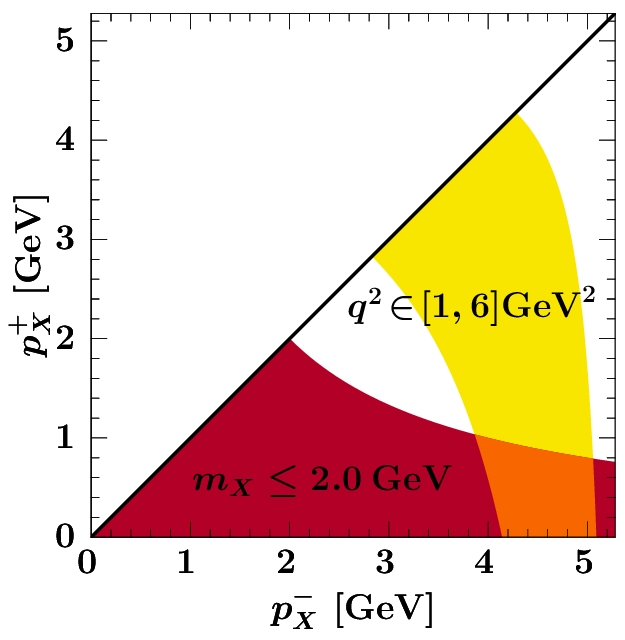}%
\caption{Phase space cuts relevant for $B\to X_s\,\ell^+\ell^-$ in the $p_X^\pm$ plane. The measurements are performed in the orange (medium) region, where the $m_X$ and $q^2$ cuts overlap and $p_X^+ \ll p_X^-$.}
\label{fig:phasespace}
\end{figure}

Without any cuts, the phase space limits on $q^2$, $\pxp$, and $z$ are
\begin{equation}\label{eq:phsp}
0 \leq \pxp \leq m_B - \sqrt{q^2} \leq m_B
\,,\qquad
-1 \leq z \leq 1
\,.\end{equation}
In the rest frame of the $B$ meson,
\begin{equation}
2m_B\,E_X = m_B^2 + m_X^2 - q^2
\,,\end{equation}
so low $q^2$ corresponds to $E_X \sim \ord{m_B}$. In conjunction with the $m_X$ cut required by the experiments we have $m_X^2 \ll E_X^2$ or equivalently $\pxp \ll p_X^-$. This is illustrated in \fig{phasespace} in the $p_X^\pm$ plane.
The measurements are done in the orange (medium) region, where the two cuts $m_X \leq 2.0\GeV$ (dark red) and $1\GeV^2\leq q^2\leq 6\GeV^2$ (light yellow) overlap. There, $p_X^-\sim m_B$ is large, while $\pxp \lesssim 1\GeV$ is small. This is precisely the kinematic region where shape function effects are important, as explained in the Introduction.

More explicitly, a cut $m_X \leq m_X^\cut$ corresponds to a $q^2$ dependent cut $\pxp \leq p_X^{+\cut}$, where
\begin{align} \label{eq:pxpcut}
 p_X^{+\,\cut} & = \frac{1}{2m_B} \biggl[
  m_B^2 + (m_X^{\rm cut})^2 - q^2
\\  & \quad
- \sqrt{\bigl(m_B^2 + (m_X^{\rm cut})^2 - q^2\bigr)^2 - 4 m_B^2(m_X^{\rm cut})^2}
  \biggr]
\nn\,.\end{align}
The $H_i$ with a cut on $m_X$ are thus given by
\begin{align}
H_i(q^2; m_X^\cut) &= \int_0^{p_X^{+\cut}}\!\!\df\pxp\, H_i(q^2, \pxp)
\,,\nn\\
H_i(q_1^2, q_2^2; m_X^\cut) &= \int_{q_1^2}^{q_2^2}\!\df q^2\, H_i(q^2; m_X^\cut)
\,,\end{align}
where the phase space limit from \eq{phsp} is implicitly understood.

\section{Split Matching and Effective Wilson Coefficients}
\label{sec:splitmatching}

After the $W$, $Z$, and $t$ are integrated out at a scale of order $m_W$, the effective weak Hamiltonian in \eq{Heff} is evolved down to the scale $m_b$, where the decay rate is calculated by evaluating the matrix elements of the operators $O_i$. In this step, the contributions from the four-quark operators $O_{1-6}$ and $O_8$ can be absorbed into effective Wilson coefficients $C_{7,9}^\mathrm{eff}(q^2)$ that are complex functions of $q^2$. In the evolution from $m_W$ down to $m_b$, $C_9(m_b)$ receives a $\ln(m_W^2/m_b^2)$ contribution from the mixing of $O_2$, which formally enhances it to $C_9(m_b) \sim \ord{1/\alpha_s}$, while numerically $\abs{C_9(m_b)} \approx C_{10}$. It is thus advantageous to separate the perturbation series above and below the scale $m_b$, such that below $m_b$ the effective Wilson coefficients can be treated as $\ord{1}$ numbers. This is achieved by the ``split matching'' procedure introduced in Ref.~\cite{Lee:2005pk} in the context of matching on to SCET.

The split matching can be thought of as first matching the effective weak Hamiltonian at a scale $\mu_0 \sim m_b$ on to a sum of effective $b\to s\,\ell^+\ell^-$ currents,
\begin{align}\label{eq:splitmatching}
&\sum_{i=1}^{10} C_i(\mu_0)\,O_i(\mu_0)
= \frac{e^2}{4\pi^2}\Bigl[C_7^\incl(q^2, \mu_0)\,J_7^\mu\,\bar\ell\gamma_\mu\ell
\\*\nn & \quad
+ C_9^\incl(q^2, \mu_0)\, J_9^\mu\,\bar\ell\gamma_\mu\ell
+ C_{10}^\incl(q^2, \mu_0)\, J_9^\mu\,\bar\ell\gamma_\mu\gamma_5\ell \Bigr]
\,,\end{align}
where
\begin{equation}\label{eq:effcurrents}
J_9^\mu = \bar{s}\,\gamma^\mu P_L b
\,,\qquad
J_7^\mu = \frac{2\,m_b}{q^2}\, \bar{s}\, \img q_\nu\sigma^{\nu\mu} P_R b\, \Big\vert_{\mu=m_b}
\,.\end{equation}
In the second step, starting from \eq{splitmatching}, the decay rate is calculated. In the local OPE treatment, the time-ordered products of the currents in \eq{effcurrents} are matched at the scale $\mu_b \sim m_b$ on to a set of local operators, whereas in SCET, the currents are matched at $\mu_b \sim m_b$ on to corresponding SCET currents, as we shall do in \sec{results}. In either case, numerically one can take $\mu_b = \mu_0$, while formally $\mu_0$ and $\mu_b$ are independent scale parameters. For example, to estimate perturbative uncertainties they can and should be varied separately.

While $J_9^\mu$ is a conserved current and thus scale-invariant, the tensor current $J_7^\mu$ has an anomalous dimension and is therefore taken to be at a fixed reference scale, $\mu = m_b$. To obtain a well-behaved
perturbative series, we use $m_b$ in the $1S$ scheme~\cite{Hoang:1998ng}, although any other
short distance $b$-quark mass could be used instead. Since both sides of \eq{splitmatching} must be $\mu_0$ independent, and the currents are (by definition) $\mu_0$ independent, the matching coefficients $C_i^\incl(q^2, \mu_0)$ are also $\mu_0$ independent to the order in perturbation theory at which the matching is performed. Hence, the decay rate calculated from \eq{splitmatching} is formally $\mu_b$ independent and one can treat the $C_i^\incl(q^2,\mu_0)$ as $\ord{1}$ when counting powers of $\alpha_s$ below the scale $\mu_b$. This also means that we have to be careful with our terminology. As far as the Wilson coefficients are concerned, we stick to the usual $B\to X_s\,\ell^+\ell^-$ counting, where, due to the formally leading $1/\alpha_s$ in $C_9(m_b)$, NNLL refers to $\ord{\alpha_s}$. On the other hand, in SCET at $\mu_b$ and below, NNLL refers to the full two-loop $\ord{\alpha_s^2}$.

Since the split matching happens at the level of currents, it captures only finite virtual corrections, which are contained in the $C_i^\incl(q^2)$, and the universal IR divergent virtual and bremsstrahlung corrections, which are described by the currents $J_{7,9}^\mu$. It does not incorporate finite bremsstrahlung corrections from operators other than $O_{7,9,10}$, which must be added explicitly. In the local OPE their effect was shown to be small, around the 1\% level~\cite{Asatryan:2002iy}\footnote{The full $z$ dependence of these corrections, which may be known from the calculations of the authors of Refs.~\cite{Asatryan:2002iy}, has not been published, and so is not known for $H_T(q^2)$ and $H_L(q^2)$ separately, but only for $H_A(q^2)$ and the sum $H_T(q^2) + H_L(q^2)$.}. In the SCET expansion they are both power and $\alpha_s$ suppressed and thus beyond the order we are working at. Similar considerations apply to electroweak corrections~\cite{Bobeth:2003at, Huber:2005ig}, which are not included here.

In Ref.~\cite{Lee:2006gs} the $C_i^\incl(q^2, \mu_0)$ are decomposed as%
\footnote{In Ref.~\cite{Lee:2006gs} the coefficients $C_i^\incl(q^2)$ are defined implicitly by absorbing all virtual contributions from $O_{1-6,8}$ into them and by requiring their $\mu_0$ independence. That definition is equivalent to the one given here. The coefficients $C_i^\mix(q^2)$ in Refs.~\cite{Lee:2005pk, Lee:2005pw} are equivalent to these except that $C_7^\mix(q^2) = (m_B/m_b) C_7^\incl(q^2)$.}
\pagebreak[4]
%
\begin{align}\label{eq:Cincl}
C_7^\incl(q^2, \mu_0) &= \C_7 + F_7(q^2) + G_7(q^2)
\,,\nn\\
C_9^\incl(q^2, \mu_0) &= \C_9 + F_9(q^2) + G_9(q^2)
\,,\nn\\
C_{10}^\incl(q^2, \mu_0) &= \C_{10}
\,,\end{align}
such that all terms on the right-hand side of \eq{Cincl} are separately
$\mu_0$ independent to the order one is working at. The explicit expressions are collected in the Appendix of Ref.~\cite{Lee:2006gs}, and we do not repeat them here. To simplify our notation we suppress the $\mu_0$ dependence in the coefficients hereafter.

The functions $F_{7,9}(q^2)$ contain the virtual contributions from $O_{1-6,8}$ and are known at NNLL order~\cite{Misiak:1992bc, Buras:1994dj, Bobeth:1999mk, Asatryan:2001de, Ghinculov:2003qd, Greub:2008cy, Gambino:2003zm} (up to small $O_{3-6}$ contributions), while the $G_{7,9}(q^2)$ contain nonperturbative $\ord{1/m_c^2}$ corrections involving the
four-quark operators~\cite{Buchalla:1997ky}. The latter can be included in this simple form, but the final results for the decay rates have to be re-expanded
so that any terms of $\ord{\alpha_s/m_c^2,\, 1/m_c^4}$ are discarded.

The coefficients $\C_{7,9,10}$ are real in the SM. They contain the dependence on the coefficients $C_{7,9,10}(\mu)$ in \eq{Heff}, i.e.
\begin{align}\label{eq:Ci_intro}
\C_7 &= C_7(\mu_0)\, \frac{\mbbar(\mu_0)}{m_b} + \dotsb
\,,\nn\\
\C_9 &= C_9(\mu_0) + \dotsb
\,,\nn\\
\C_{10} &\equiv C_{10}
\,,\end{align}
which are sensitive to new physics.
On the other hand, the functions $F_i(q^2)$ and $G_i(q^2)$ are dominated by contributions from $O_{1,2}$ and thus are expected to be insensitive to new physics. Hence,
the approach advocated in Ref.~\cite{Lee:2006gs} to search for new physics is to assume the SM everywhere and treat $\C_{7,9,10}$ as three unknown real parameters to be extracted from data; it was shown that $H_L(1,6)$, $H_T(1,6)$, $H_A(1,3.5)$, and $H_A(3.5,6)$ are sufficient for this purpose. This strategy has the advantage that the number of parameters is kept to a minimum and thus the sensitivity to new physics can be maximized. In addition, there is no dependence on a specific new physics model. New physics contributions will show up as inconsistencies between the extracted values of $\C_{7,9,10}$ and their calculated SM values, or between overconstraining measurements (similar to the usual approach to overconstrain the Cabibbo-Kobayashi-Maskawa matrix).

\bigskip
\section{Results}
\label{sec:results}

In this section, we present our results for the three observables $H_T$, $H_A$, and $H_L$ defined in \eq{d3Gamma} in the SCET region, $\pxp \ll p_X^-$. We write their structure functions in \eq{HTAL} as
\begin{equation}
W_i(q^2, p_X) = W_i^{[0]}(q^2, \pxp) + W_i^{[1]}(q^2, \pxp) + \dotsb
\,,\end{equation}
where $i=T,A,L$ and the superscript $[n]$ denotes the order $(\lqcd/m_b)^n$ in the power expansion. The leading-order $W_i^{[0]}(q^2, \pxp)$, involving the leading shape function, are discussed next, while the $W_i^{[1]}(q^2, \pxp)$, containing the subleading shape functions, are discussed in \subsec{NLO}.

\subsection{Leading order}
\label{subsec:LO}

The leading-order structure functions factorize as
\begin{align} \label{eq:W0fact}
&W_i^{[0]}(q^2,\pxp) = h_i^{[0]}(q^2,\pxp,\mu_i)
\nn\\&\quad
\times\int \!\df\w\, p^-J(p^-\w,\mu_i)\, S(\pxp - \w, \mu_i)
\,,\end{align}
where $p^- = m_b - q_- = m_b - q^2/(m_B-\pxp)$ is the partonic light-cone momentum. The integration limits here and below are implicit in the support of the functions, which are nonzero only if their first argument is positive. We shall discuss each ingredient in \eq{W0fact} in turn.

The hard functions $h_i^{[0]}(q^2,\pxp, \mu_b)$ are different for each structure function. To obtain them, we start by matching the QCD currents in \eq{effcurrents} at the hard scale $\mu_b \sim m_b$ on to corresponding SCET currents,
\begin{align}\label{eq:SCETmatching}
J_9^\mu &= \sum_{i=1,2,3} c^9_i(p^-,\mu_b)\,\bar\chi_n\,\Gamma_{9,i}^\mu\,\cH^n_v
\,,\nn\\
J_7^\mu &= \frac{2m_b}{q^2}\,\sum_{i=1,2} c^7_i(p^-,\mu_b)\,\bar\chi_n\,\Gamma^\mu_{7,i}\,\cH^n_v
\,,\end{align}
where $\chi_n = W_n^\dagger\xi_n$ and $\cH^n_v = Y^\dagger_n b_v$ are the standard collinear and heavy-quark fields in SCET, and $p^-$ corresponds to the large momentum label on the collinear quark field. We choose a slightly different set of minimal Dirac structures than usual,
\begin{align}
\Gamma_{9,i}^\mu &= P_R\,\bigl\{\gamma^\mu, v^\mu, q^\mu \bigr\}
\,,\nn\\
\Gamma_{7,i}^\mu &= P_R\,\bigl\{\img q_\nu\sigma^{\nu\mu}, q_\nu (q^\nu v^\mu-q^\mu v^\nu) \bigr\}
\,.\end{align}
The reason to use $q^\mu$ instead of $n^\mu$ for $\Gamma^\mu_{9,3}$ is that it makes explicit the constraint from lepton current conservation, which implies that for massless leptons only two coefficients contribute to the rate. For $\Gamma^\mu_{7,i}$ there are only two independent coefficients from the start, because $q_\mu J_7^\mu = 0$ provides an additional constraint.

The matching for general currents to $\ord{\alpha_s}$ was carried out in Ref.~\cite{Bauer:2000yr}. For the vector current, the two-loop $\ord{\alpha_s^2}$ matching has become available only recently, through the work of several groups~\cite{Bonciani:2008wf, Asatrian:2008uk, Beneke:2008ei, Bell:2008ws}. We find
\begin{widetext}
\begin{align} \label{eq:c9i}
c^9_1(p^-, \mu_b)
&= 1 - \frac{\alpha_s(\mu_b)}{2\pi}\,C_F \biggl[
  \ln^2\!\frac{\mu_b}{p^-} + \frac{5}{2}\ln\frac{\mu_b}{p^-} + {\rm Li}_2\Bigl(1-\frac{p^-}{m_b}\Bigr)
  + \frac{1}{2}\ln\frac{p^-}{m_b}\,\Bigl(\frac{m_b}{m_b - p^-} + 2\Bigr) + \frac{\pi^2}{24} + 3 \biggr]
\nn\\* & \quad
  + \frac{\alpha_s^2(\mu_b)}{16\pi^2}\,C_1^{(2)}\Bigl(\frac{p^-}{m_b}, \mu_b\Bigr)
\,,\nn\\*
c^9_2(p^-, \mu_b)
&= \frac{\alpha_s(\mu_b)}{2\pi}\,C_F\,\ln\frac{p^-}{m_b}\,\frac{m_b}{m_b - p^-}
  + \frac{\alpha_s^2(\mu_b)}{16\pi^2}\,\biggl[C_2^{(2)}\Bigl(\frac{p^-}{m_b}, \mu_b\Bigr)
  + \frac{2m_b}{p^-}\, C_3^{(2)}\Bigl(\frac{p^-}{m_b}, \mu_b\Bigr)\biggr]
\,,\nn\\
c^9_3(p^-, \mu_b)
&= \frac{\alpha_s(\mu_b)}{2\pi}\,C_F \frac{1}{m_b- p^-}\biggl[
   \ln\frac{p^-}{m_b}\,\Bigr(\frac{m_b}{m_b- p^-} - 2\Bigl) + 1 \biggr]
  - \frac{\alpha_s^2(\mu_b)}{16\pi^2}\,\frac{2}{p^-}\,C_3^{(2)}\Bigl(\frac{p^-}{m_b},\mu_b\Bigr)
\,,\end{align}
where the two-loop functions $C_i^{(2)}(u)$ can be found in Ref.~\cite{Bell:2008ws}, and as indicated they have to be evaluated at $\mu = \mu_b$. For the tensor current, we find
\begin{align} \label{eq:c7i}
c^7_1(p^-,\mu_b)
&= 1 - \frac{\alpha_s(\mu_b)}{2\pi}\,C_F \biggl[
  \ln^2\!\frac{\mu_b}{p^-} + \frac{5}{2}\ln\frac{\mu_b}{p^-} + {\rm Li}_2\Bigl(1-\frac{p^-}{m_b}\Bigr)
  + \frac{3}{2}\ln\frac{p^-}{m_b} + \frac{\pi^2}{24} + 3 \biggr]
\,,\nn\\
c^7_2(p^-,\mu_b)
&= -\frac{\alpha_s(\mu_b)}{2\pi}\,C_F\,\ln\frac{p^-}{m_b}\,\frac{2}{m_b - p^-}
\,.\end{align}
The $\ord{\alpha_s^2}$ corrections for the tensor current are not fully known at present, but since two-loop calculations for the vector current exist, the equivalent two-loop calculation for the tensor current should be feasible. From the two-loop
computation of the $\abs{C_7}^2$ terms in the $b\to s\gamma$ rate~\cite{Melnikov:2005bx, Blokland:2005uk}, one can obtain the $\ord{\alpha_s^2}$ contribution to $c_1^7$ at the point $p^- = m_b$~\cite{Ali:2007sj, Ligeti:2008ac}. For the vector current, the $\alpha_s^2$ corrections to $c^9_i(p^-,\mu_b)$ for small $q^2$ or large $p^-$ are to good approximation given by a constant shift. Assuming a similar behavior for the tensor current, we can obtain an approximate result for $c_1^7$ at $\ord{\alpha_s^2}$ in the low $q^2$ region,
\begin{equation} \label{eq:c71approx}
\tilde c^7_1(p^-,\mu_b) = c_1^7(p^-,\mu_b) + \frac{1}{2}\bigl[h_s(m_b, \mu_b) - c_1^7(m_b,\mu_b)^2\bigr]
\,,\end{equation}
where $c^7_1$ is the result to $\ord{\alpha_s}$ in \eq{c7i}, and $h_s(m_b,\mu_b)$ is given to $\ord{\alpha_s^2}$ in Eq.~(A4) of Ref~\cite{Ligeti:2008ac}.

The hard functions $h_i^{[0]}$ are now computed by substituting the currents in \eq{SCETmatching} together with their prefactors from \eq{splitmatching} into \eq{Wmunu} for the hadronic tensor and factorizing out the matching coefficients times the trace of their Dirac structures. One then obtains $W_{\mu\nu}^{[0]} = h_{\mu\nu}^{[0]}\, J\otimes S$, with (writing for simplicity $c^{10}_i \equiv c^9_i$ and $\Gamma_{10,i} \equiv \Gamma_{9,i}$)
\begin{equation}
h^{[0]\,\mu\nu} =\!\! \sum_{a,b=7,9,10}\!\!\! C_a^{\incl *}\,C_b^{\incl}\,\sum_{i,j=1,2}\!\! c_i^a c_j^b\, \mathrm{Tr}\Bigl[\frac{1+\vslash}{2}\,\overline\Gamma^\mu_{a,i}\,\frac{\nslash}{4}\,\Gamma^\nu_{b,j}\Bigr]
\,.\end{equation}
The remaining matrix element gives the convolution of jet and shape function, $J\otimes S$.

Taking the traces and the appropriate linear combinations from \eq{WTAL}, we find~\cite{Lee:2005pk}
\begin{align} \label{eq:hi0}
h_T^{[0]}(q^2,\pxp, \mu_b)
&= \ABs{C_9^\incl(q^2)\,c^9_1(p^-, \mu_b) + \frac{2m_b}{q_-}\, C_7^\incl(q^2)\, c^7_1(p^-, \mu_b)}^2
     + \C_{10}^2\,\bigl[c^9_1(p^-,\mu_b)\bigr]^2
\,,\nn \\
h_A^{[0]}(q^2,\pxp, \mu_b) &= 2\,\C_{10}\, c^9_1(p^-, \mu_b) \Re\Bigl[
C_9^\incl(q^2)\, c^9_1(p^-, \mu_b) + \frac{2m_b}{q_-}\,C_7^\incl(q^2)\, c^7_1(p^-, \mu_b)\Bigr]
\,,\nn \\
h_L^{[0]}(q^2,\pxp, \mu_b)
&= \ABS{C_9^\incl(q^2)\,\Bigl[q_+\,c^9_1(p^-, \mu_b) + \frac{q_+ - q_-}{2}\,c^9_2(p^-, \mu_b)\Bigr]
        + 2m_b\, C_7^\incl(q^2) \Bigl[c^7_1(p^-, \mu_b) + \frac{q_+-q_-}{2}\,c^7_2(p^-, \mu_b)\Bigr]}^2
\nn\\ & \quad
   + \C_{10}^2\, \Bigl[q_+\, c^9_1(p^-, \mu_b) + \frac{q_+ - q_-}{2}\,c^9_2(p^-, \mu_b) \Bigr]^2
\,.\end{align}
\end{widetext}
To evolve the coefficients from the hard scale $\mu_b\sim m_b$ to the intermediate scale $\mu_i\sim m_X\sim \sqrt{\lqcd m_b}$, we use
\begin{equation}\label{eq:hardevolution}
h_i^{[0]}(q^2,\pxp, \mu_i) = h_i^{[0]}(q^2,\pxp, \mu_b)\, U_H(p^-,\mu_b,\mu_i)
\,,\end{equation}
where the hard evolution factor~\cite{Bauer:2000yr} sums logarithms between the scales $\mu_b$ and $\mu_i$ and is known at NNLL.

Next, we consider the convolution of jet and shape function in \eq{W0fact}.
The jet function $J(p^-\w, \mu_i)$ contains perturbative physics at the intermediate jet scale $\mu_i\sim m_X$, and is known at $\ord{\alpha_s}$~\cite{Bauer:2003pi, Bosch:2004th} and $\ord{\alpha_s^2}$~\cite{Becher:2006qw}.

The leading shape function $S(\w, \mu)$ is defined as%
\footnote{We use a different normalization of the $\ket{B}$ state from that in Ref.~\cite{Ligeti:2008ac}.}
\begin{equation}\label{eq:S_def}
S(\w, \mu) = \frac{1}{2m_B}\,\mae{B}{O_0(\w, \mu)}{B} \equiv \vevB{O_0(\w, \mu)}\,,
\end{equation}
where
\begin{equation}
O_0(\w, \mu) = \bar b_v\,\delta(\img D_+ - \delta+\w) b_v
\,.\end{equation}
Here, $b_v$ is the HQET $b$ quark field, $\img D^\mu$ is an ultrasoft covariant derivative, and $\delta = m_B - m_b$, so $S(\w)$ has support for $\w > 0$. We use the full QCD $B$ meson state $\ket{B}$ in \eq{S_def}, which automatically absorbs into $S(\w, \mu)$ all subleading shape functions that would otherwise arise from time-ordered products of $O_0(\w, \mu)$ with the power corrections in the HQET Lagrangian.

The shape function contains both nonperturbative and perturbative physics. A method to combine all available perturbative and nonperturbative information was developed recently in Ref.~\cite{Ligeti:2008ac}. To do so, the shape function at the scale $\mu_i$ is written as
\begin{align} \label{eq:S}
S(\w,\mu_i) &= \int\!\df\w'\int\!\df k\, U_S(\w-\w', \mu_i, \mu_\Lambda)
\nn\\ & \quad\times
\hC_0(\w'-k, \mu_\Lambda)\, \hF(k)
,\end{align}
where the hats indicate that the quantities are given in a renormalon-free short distance scheme.
The function $\hC_0$ is perturbatively calculable at the soft scale $\mu_\Lambda$, and is known at $\ord{\alpha_s}$~\cite{Bauer:2003pi, Bosch:2004th} and $\ord{\alpha_s^2}$~\cite{Becher:2005pd}.
The soft evolution factor $U_S$~\cite{Balzereit:1998yf, Bosch:2004th, Fleming:2007xt} sums logarithms between the soft scale $\mu_\Lambda$ and the jet scale $\mu_i$.
Finally, $\hF(k)$ is a fully nonperturbative function, which can be constrained by data from $B\to X_s\gamma$, $B\to X_u\ell\bar\nu$ and $B\to X_c\ell\bar\nu$~\cite{Ligeti:2008ac}.

Combining the convolutions in \eqs{S}{W0fact}, we define the perturbative function $\hP(p^-, \pxp, \mu_i)$ by
\begin{align} \label{eq:P_def}
\hP(p^-, \pxp, \mu_i) &= \int\!\df\w\int\!\df\w'\ p^- J[p^-(\pxp - \w), \mu_i]
\nn\\
& \quad\times U_S(\w - \w', \mu_i, \mu_\Lambda)\, \hC_0(\w', \mu_\Lambda)
\nn\\
&= \delta(k) + \ord{\alpha_s}
\,,\end{align}
and combining this with \eq{hardevolution} we obtain
\begin{align} \label{eq:W0final}
W_i^{[0]}(q^2,\pxp) &= h_i^{[0]}(q^2,\pxp,\mu_b)\,U_H(p^-,\mu_b,\mu_i)
\nn\\&\quad
\times\int \!\df k\, \hP(p^-, \pxp-k, \mu_i)\,\hF(k)
\,.\end{align}
With \eq{hi0} and the matching coefficients in \eqs{c9i}{c71approx} we have an approximate $\ord{\alpha_s^2}$ result for $h_{T,A}^{[0]}$, which do not depend on $c_2^7$. While $h_L^{[0]}$ depends on $c_2^7$, it has no soft photon pole and is thus completely dominated by the vector current contributions, which are known at $\ord{\alpha_s^2}$. An explicit expression for $\hP(p^-, \pxp, \mu_i)$ to $\ord{\alpha_s^2}$ with NNLL summation, in any short distance scheme for the $b$-quark mass and kinetic-energy matrix element, has been derived in Ref.~\cite{Ligeti:2008ac}. An explicit NNLL expression for $U_H$ can be found there as well.
Hence, approximate NNLL $\ord{\alpha_s^2}$ results are available at leading order in the SCET power expansion for all three observables $H_{T,A,L}(q^2,\pxp)$.

\subsection{Subleading order}
\label{subsec:NLO}

At tree level and $\ord{\lqcd/m_b}$ six additional subleading shape functions enter in the description of $B\to X_u\ell\bar\nu$ and $B\to X_s\gamma$~\cite{Bauer:2001mh, Leibovich:2002ys, Bauer:2002yu, Burrell:2003cf, Mannel:2004as, Lee:2004ja, Bosch:2004cb, Beneke:2004in, Tackmann:2005ub}, and will also contribute to $B\to X_s\,\ell^+\ell^-$. We refer to these as the primary subleading shape functions. The analog of the factorization theorem \eq{W0fact} at $\ord{\lqcd/m_b}$ was worked out explicitly in Ref.~\cite{Lee:2004ja}. At $\ord{\alpha_s\lqcd/m_b}$ an even larger number of additional shape functions appears~\cite{Lee:2004ja, Beneke:2004in, Trott:2005vw}. The split matching relies on the fact that for $O_{7,9,10}$ we can treat $q^2$ as $\ord{1}$ in the SCET expansion. If subleading contributions from other operators are considered, it can be necessary to count $q^2$ as parametrically small and to treat the photon as collinear particle. In this case there will be additional four-quark operators with collinear quarks coupling to the collinear photon, giving rise to additional subleading four-quark shape functions~\cite{Lee:2005pk, Lee:2006wn}.
We shall restrict our discussion to tree level and the primary subleading shape functions.

When we consider the $\ord{\lqcd/m_b}$ power corrections, the split matching is important for two reasons. First, it is convenient, because it allows us to think only about the two currents in \eq{effcurrents}. This implies that the factorization in Ref.~\cite{Lee:2004ja} also applies to our case, and a large part of the results can be reused. More importantly, it provides us with a consistent way to work at tree level at the scale $\mu_b$ and below and neglect $\ord{\alpha_s\lqcd/m_b}$ loop corrections in SCET, while at the same time keeping the full $\alpha_s$ corrections to the effective Wilson coefficients from scales $\mu_0$ and above, even when they multiply subleading shape functions. In this way, we can avoid artificially large power corrections that arise simply from having to use different Wilson coefficients at $\ord{\lqcd/m_b}$, and can instead use the same Wilson coefficients at each order in the power counting.

The calculation proceeds along the same lines as in the previous section, though here there are two sources of subleading corrections. First, the matching in \eq{SCETmatching} now has to include subleading SCET currents. Secondly, when the time-ordered products are evaluated, there will be corrections from higher-order terms in the SCET Lagrangian. Alternatively, working at tree level, we can directly match the time-ordered products of the effective currents on to the subleading shape function operators as in Ref.~\cite{Tackmann:2005ub}. Of course, both approaches give the same results.

The operators that arise from subleading SCET currents are
\begin{align}
O_{1}^\mu(\w) &= \frac{1}{2}\, \bar b_v\, \bigl\{\img D^\mu, \delta(\img D_+ - \delta + \w) \bigr\}\, b_v
\,, \nn \\
P_2(\w)
&= \frac{\img}{2}\,\epsilon_{\perp\mu\nu}\, \bar b_v\, \bigl[\img D^\mu, \delta(\img D_+ - \delta + \w) \bigr]\,
   \gamma_T^\nu \gamma_5\, b_v
\,.\end{align}
They come with the same jet function as the leading-order shape function. The contribution from $O_1^\mu(\w)$ can be rewritten in terms of the leading-order result as
\begin{align}
&\int\!\df\w\, p^- J(p^-(\pxp - \w), \mu_i)\, n_\mu\vevB{O_1^\mu(\w)}
\nn\\
&= \int\!\df\w\, p^- J(p^-(\pxp - \w), \mu_i)\, (\delta - \w)S(\w)
\nn\\
&= (\delta-\pxp)\hF(\pxp) + \ord{\alpha_s}
\,,\end{align}
while $P_2(\w)$ gives rise to a new subleading shape function,
\begin{equation}
\int\!\df\w\, p^- J(p^-(\pxp - \w), \mu_i)\, \vevB{P_2(\w)} = F_2(\pxp) + \ord{\alpha_s}
\,.\end{equation}

The operators that are due to higher-order terms in the SCET Lagrangian are
\begin{widetext}
\begin{align}
O_3(\w_1, \w_2)
&= \bar b_v\, \delta(\img D_+ - \delta + \w_1)\,(\img D_\perp)^2\, \delta(\img D_+ - \delta + \w_2)\, b_v
\,,\nn \\
P_4^{\mu}(\w_1,\w_2)
&= \frac{1}{2}\, \bar b_v\,\delta(\img D_+ - \delta + \w_1)\, g\epsilon_{\perp\nu\lambda}G_\perp^{\nu\lambda}\, \delta(\img D_+ - \delta + \w_2)\, \gamma_T^\mu \gamma_5\,  b_v
\,, \nn \\
O_{5s}^{\mu\nu}(\w_1,\w_2,\w_3)
&= \bigl[\bar b_v T^A\, \delta(\img D_+ - \delta + \w_1)\, \gamma^\mu P_L s_{us}^\bn\bigr]
  \delta(\img\partial_+ - \delta + \w_2)
\bigl[\bar s_{us}^\bn \gamma^\nu P_L\, \delta(\img D_+ - \delta + \w_3)\,T^A b_v \bigr]
\,,\end{align}
where $\epsilon_\perp^{\mu\nu} = \epsilon^{\mu\nu\alpha\beta} n_\alpha \bn_\beta/2$ (with $\epsilon_{0123} = 1$), $\img g G_\perp^{\mu\nu}=[\img D_\perp^\mu, \img D_\perp^\nu]$, and $s_{us}^\bn=(\bnslash\nslash/4)\, s_{us}$ is an ultrasoft $s$ quark field. These operators are associated with new jet functions that are known only at tree level. Combining their $B$ matrix elements with their jet functions, we define
\begin{align}
F_3(\pxp) &= \int\!\df\w_1\,\df\w_2\,\biggl[\frac{\delta(\pxp-\w_1)}{\pxp-\w_2} + \frac{\delta(\pxp-\w_2)}{\pxp-\w_1}\biggr] \vevB{O_3(\w_1, \w_2)}
\,,\nn\\
F_4(\pxp) &= \int\!\df\w_1\,\df\w_2\,\biggl[\frac{\delta(\pxp-\w_1)}{\pxp-\w_2} + \frac{\delta(\pxp-\w_2)}{\pxp-\w_1}\biggr] n_\mu\vevB{P_4^\mu(\w_1,\w_2)}
\,,\nn\\
F_5^s(\pxp) &=\int\!\df\w_1\,\df\w_2\,\df\w_3\, \frac{4\pi\alpha_s}{2\pi\img}\biggl[
\prod_{j=1,2,3}\frac{1}{\pxp-\w_j-\img\e} -\prod_{j=1,2,3}\frac{1}{\pxp-\w_j+\img\e} \biggr]
n_\mu n_\nu\vevB{O_{5s}^{\mu\nu}(\w_1,\w_2,\w_3)}
\,,\nn\\
F_6^s(\pxp) &= \int\!\df\w_1\,\df\w_2\,\df\w_3\, \frac{4\pi\alpha_s}{2\pi\img}\biggl[
\prod_{j=1,2,3}\frac{1}{\pxp-\w_j-\img\e} -\prod_{j=1,2,3}\frac{1}{\pxp-\w_j+\img\e} \biggr]
(g^\perp_{\mu\nu} + \img\epsilon^\perp_{\mu\nu})\,\vevB{O_{5s}^{\mu\nu}(\w_1,\w_2,\w_3)}
\,,\end{align}
which correspond to the functions defined in Ref.~\cite{Lee:2004ja}.
There are also operators $P_1^\mu$, $O_2$, $P_3$, and $O_4^\mu$, obtained from the above by interchanging the Dirac structure $1 \leftrightarrow \gamma^\mu_T\gamma_5$. They do not contribute because their matrix elements between $B$ meson states vanish as a result of parity and/or time-reversal invariance.

With the above definitions, we find the following $\ord{\lqcd/m_b}$ corrections to the structure functions:
\begin{align}\label{eq:W1result}
W_T^{[1]}(q^2,\pxp)
&= -\bigl[\abs{C_9^\incl(q^2)}^2 + \C_{10}^2\bigr] \biggl[\frac{F_1(\pxp)}{m_b} + \frac{F_T(\pxp)}{p^-}\biggr]
+ \frac{4m_b^2}{q_-^2}\,\abs{C_7^\incl(q^2)}^2 \biggl[\frac{F_1(\pxp)}{m_b} - \frac{F_T(\pxp)}{p^-}\biggr]
\nn\\ & \quad
- \frac{4m_b}{q_-}\Re\bigl[C_9^\incl(q^2) C_7^{\incl*}(q^2)\bigr]\frac{F_T(\pxp)}{p^-}
\,,\nn\\
W_A^{[1]}(q^2,\pxp)
&=-2\,\C_{10} \Re\biggl\{
C_9^\incl(q^2) \biggl[\frac{F_1(\pxp)}{m_b} + \frac{F_T(\pxp)}{p^-}\biggr]
+ \frac{2m_b}{q_-}\,C_7^\incl(q^2) \frac{F_T(\pxp)}{p^-} \biggr\}
\,,\nn\\
W_L^{[1]}(q^2,\pxp)
&= \bigl[\abs{C_9^\incl(q^2)}^2 + \C_{10}^2\bigr]
\biggl[\frac{q_+^2 F_1(\pxp)}{m_b} - \frac{q_+^2F_L(\pxp) + 2q^2F_2(\pxp)}{p^-}\biggr]
\nn\\ & \quad
- 4m_b^2\,\abs{C_7^\incl(q^2)}^2 \biggl[\frac{F_1(\pxp)}{m_b} + \frac{F_L(\pxp) + 2F_2(\pxp)}{p^-}\biggr]
\nn\\ & \quad
- 4m_b\,\Re\bigl[C_9^\incl(q^2) C_7^{\incl*}(q^2)\bigr] \frac{q_+ F_L(\pxp) + (q_+ + q_-)F_2(\pxp)}{p^-}
\,,\end{align}
\pagebreak[4]
\end{widetext}
where we have used the abbreviations
\begin{align}\label{eq:Fi_eff}
F_1(\pxp) &= (\pxp - \delta)F(\pxp) + F_2(\pxp)
\,,\nn\\
F_T(\pxp) &= F_3(\pxp) - F_4(\pxp) + 2F_5^s(\pxp)
\,,\nn\\
F_L(\pxp) &= F_3(\pxp) + F_4(\pxp) - 2F_6^s(\pxp)
\,.\end{align}
Note that in $W_T^{[1]}$ and $W_A^{[1]}$ only two different combinations of subleading shape functions appear, a property which can be exploited to construct particular combinations of observables in $B\to X_u\ell\bar\nu$ and $B\to X_s\ell^+\ell^-$ for which the subleading shape functions drop out~\cite{Lee:2008vs}.

\section{\boldmath $m_X$-cut Effects At Subleading Order}
\label{sec:SSFeffects}

In this section, we briefly investigate the numerical impact of the power corrections in \eq{W1result} on the different observables, using the input values collected in Table~\ref{tab:num}. A detailed numerical analysis of the $m_X$ cut effects including an estimation of uncertainties is beyond the scope of this paper and is relegated to a dedicated publication~\cite{toappear}.

To obtain expressions for the leading and subleading shape functions, we follow the construction in Ref.~\cite{Ligeti:2008ac}. We give only a few relevant formulas here, and refer the reader to Ref.~\cite{Ligeti:2008ac} for further details. The nonperturbative function $\hF(k)$ entering the leading-order result \eq{W0final} can be expanded as
\begin{equation} \label{eq:expdef}
\hF(\lambda\,x) = \frac{1}{\lambda}\,\biggl[ \sum_{n = 0}^\infty c_n\, f_n(x) \biggr]^2
\,,\end{equation}
where $\lambda\sim\lqcd$ is a free parameter and $f_n(x)$ form a complete set of orthonormal functions on $[0,\infty)$. We use the default value $\lambda = 0.8\GeV$ and $f_n(x)$ from Eq.~$(48)$ of Ref.~\cite{Ligeti:2008ac}.

\begin{table}[t]
\tabcolsep 6pt
\begin{tabular}{c|c}
\hline\hline
Parameter & Value \\
\hline
$m_B$ & $5.279\GeV$ \\
$m_b\equiv m_b^{1S}$~\cite{Barberio:2008fa} & $4.70\GeV$ \\
\hline
$\lambda_1\equiv\lambda_1^\img$~\cite{Ligeti:2008ac} & $-0.32\GeV^2 $ \\
$\lambda_2$ & $0.12\GeV^2 $ \\
$\rho_2$~\cite{Barberio:2008fa} & $-0.065\GeV^3$ \\
\hline
$\C_7$ & $-0.2611$ \\
$\C_9$ & $4.207$ \\
$\C_{10}$ & $-4.175$ \\
\hline\hline
\end{tabular}
\caption{Central values of input parameters.}
\label{tab:num}
\end{table}

Since our main interest is in the corrections from subleading shape functions, we use a fixed model for $\hF(k)$, obtained by truncating the series in \eq{expdef} at $n\leq 2$. For a given value of $\lambda$, the remaining coefficients $c_{0,1,2}$ are determined by the $0$th, $1$st, and $2$nd moments of $\hF(k)$,
\begin{align} \label{eq:Fhatmoments}
 &  \int\! \df k\, \hF(k) =  1
  \,,\nn\\
 &  \int\! \df k\, k\, \hF(k) = m_B - m_b^{1S}
  \,,\nn\\
 &  \int\! \df k\, k^2\, \hF(k) = -\frac{\lambda_1^\img}{3} + (m_B - m_b^{1S})^2
\,,\end{align}
with $m_b^{1S}$ given in the $1S$ scheme and $\lambda_1^\mathrm{i}$ in the ``invisible'' scheme~\cite{Ligeti:2008ac}.

\begin{figure}[t]
\includegraphics[width=\columnwidth]{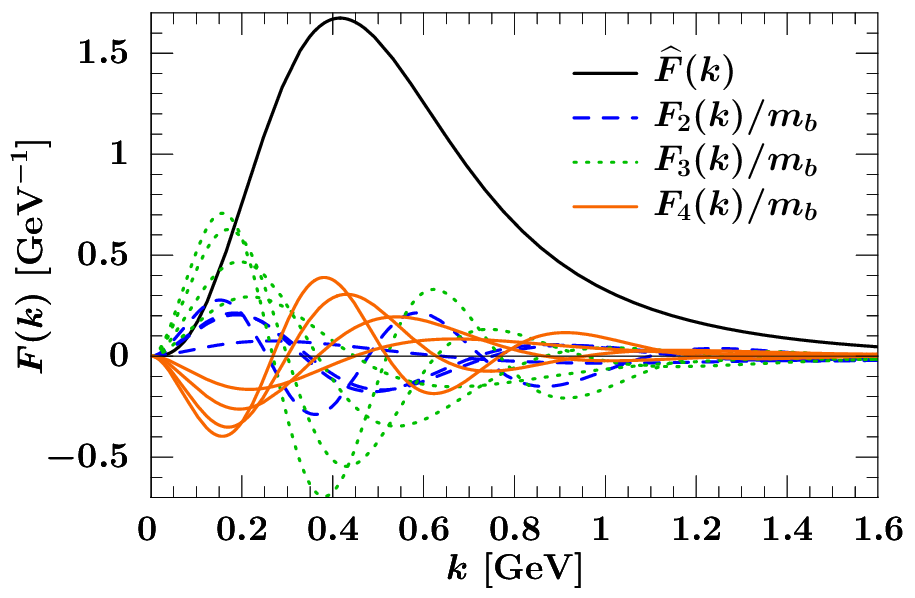}%
\caption{Model functions used for the subleading shape functions $F_2(k)$ (dashed blue), $F_3(k)$ (dotted green), and $F_4(k)$ (solid orange). The black solid line shows the model used for the leading-order function $\hF(k)$.}
\label{fig:SSFmodels}
\end{figure}

\begin{figure*}[t]
\includegraphics[width=\columnwidth]{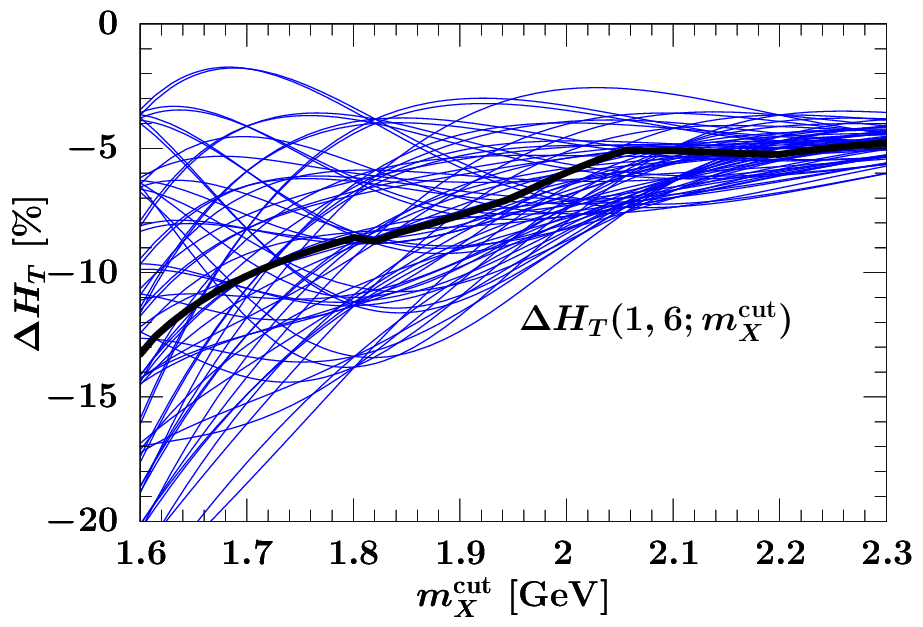}%
\hspace{\columnsep}%
\includegraphics[width=\columnwidth]{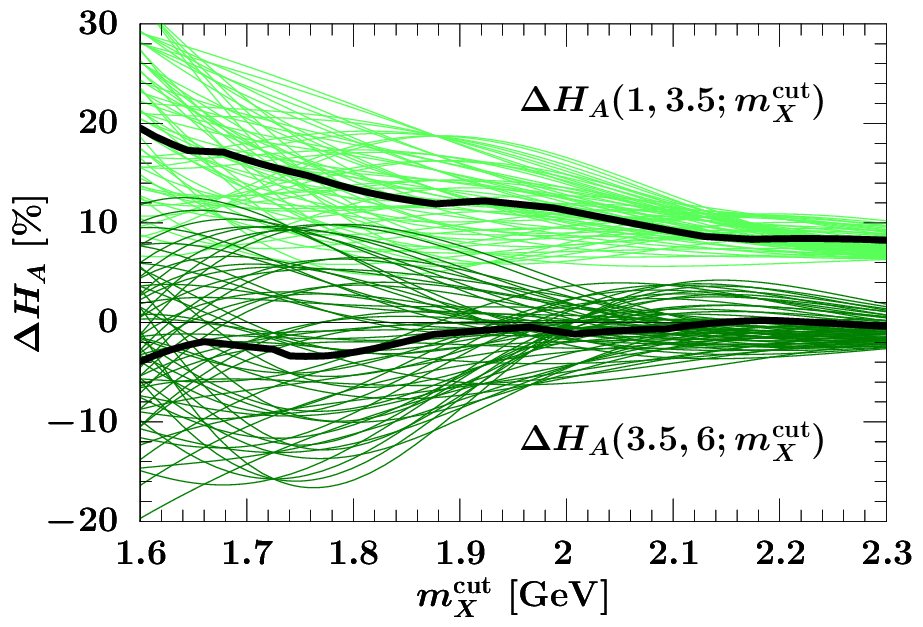}%
\\
\includegraphics[width=\columnwidth]{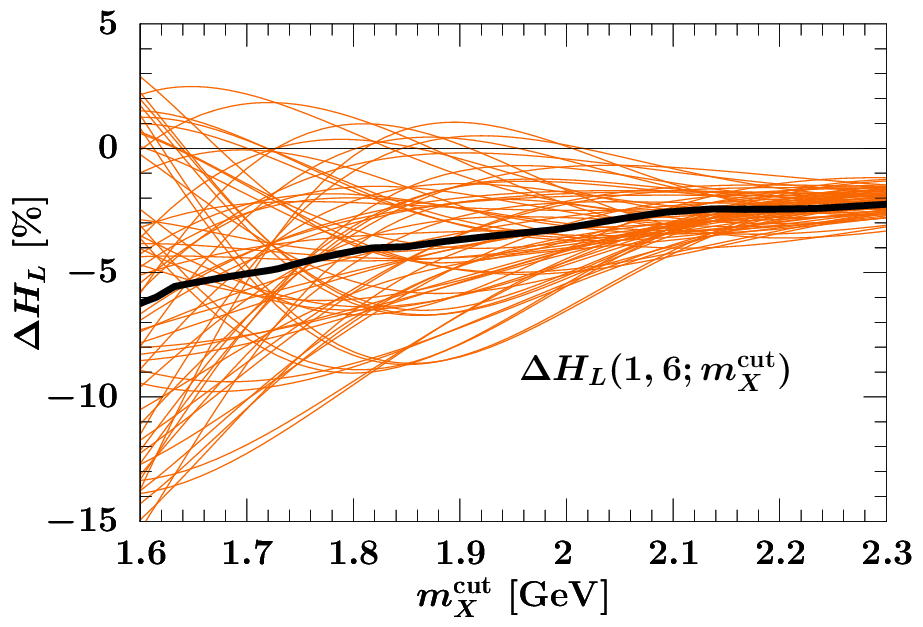}%
\hspace{\columnsep}%
\includegraphics[width=\columnwidth]{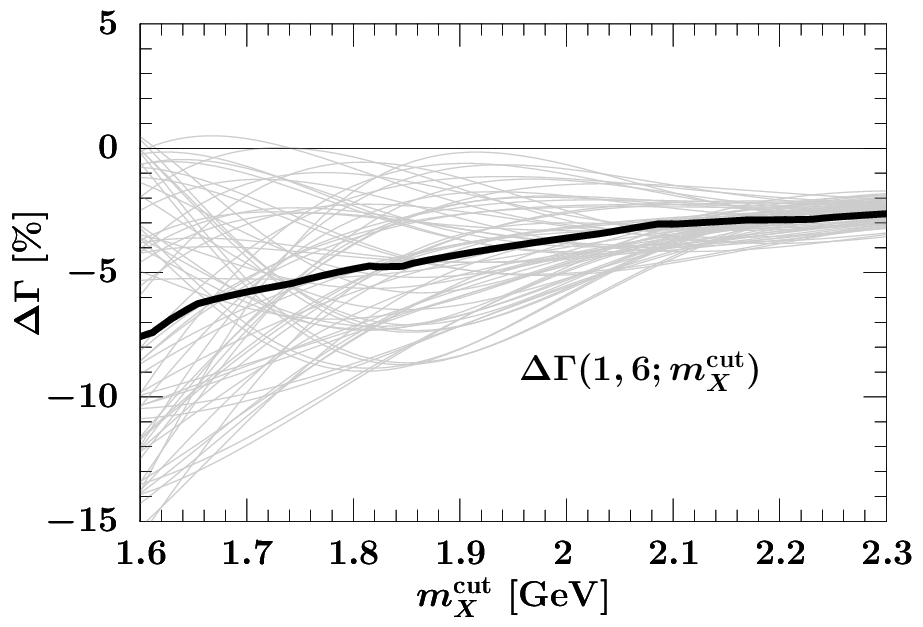}%
\caption{Relative corrections due to subleading shape functions as function of $m_X^\cut$ for $H_T(1,6;m_X^\cut)$ (top left), $H_A(1,3.5;m_X^\cut)$ and $H_A(3.5,6;m_X^\cut)$ (top right), $H_L(1,6;m_X^\cut)$ (bottom left), and $\Gamma(1,6;m_X^\cut)$ (bottom right). The thin lines show the result of using the different subleading shape function models from \fig{SSFmodels}. The thick black line in each case shows the center of the thin curves.}
\label{fig:Hicorr}
\end{figure*}

Very little is known about the subleading shape functions. Since the flavor of the light quark in the operator $O_{5s}$ does not match the flavor of the spectator quark in the $B$ meson, we expect the functions $F_{5,6}^s(\pxp)$ to give only small corrections. Furthermore, since they arise only in combination with $F_{3,4}(\pxp)$ as in \eq{Fi_eff}, we can assume that any small effect they may have will likely be washed out by the uncertainties in $F_{3,4}(\pxp)$. We therefore set $F_{5,6}(\pxp)$ to zero in our numerical analysis.
The first moments of the remaining functions are
\begin{align} \label{eq:sublmoments}
\int\!\df k\,F_{2,3,4}(k) &= 0
\,,\nn\\
\int\!\df k\,k\,F_{2,3,4}(k) &= \{-\lambda_2, 2\lambda_1/3, \lambda_2\}
\,,\nn\\
\int\!\df k\,(k - \delta)^2\,F_{2,3,4}(k) &= \{\rho_2, 0, 0\}
\,.\end{align}
For $F_{2,3,4}(k)$ we use a construction similar to \eq{expdef},
\begin{equation}
F_i(\lambda\,x) = \mp\frac\df{\df x}\, \biggl[ \sum_{n=0}^\infty  d^i_n\, f_n(x) \biggr]^2
\,,\end{equation}
which automatically incorporates the vanishing $0$th moment. The overall sign is determined by the sign of the first moment. To obtain a range of models for each function we consider two cases, $d^i_{0,1}\neq 0$ and $d^i_{1,2}\neq 0$, with all other coefficients set to zero. For each case, there are two solutions to the moment constraints from \eq{sublmoments}, providing us with a total of four reasonably different models for each function, which are shown in \fig{SSFmodels}. When combined, these give $64$ different sets of models for the subleading shape functions, which we use to illustrate their effects. We stress that the spread in the results obtained from these models should not be interpreted as a rigorous theoretical error, but merely as an indication of the rough size of the uncertainty expected from the unknown form of the subleading shape functions. A more detailed analysis will be presented in Ref.~\cite{toappear}.

To illustrate the effect of the power corrections, we consider their relative corrections to the lowest-order result,
\begin{equation}
\Delta H_i(q_1^2, q_2^2; m_X^\cut)
= \frac{H_i^{[1]}(q_1^2, q_2^2; m_X^\cut)}{H_i^{[0]}(q_1^2, q_2^2; m_X^\cut)}
\,.\end{equation}
Here, $H_i = \{H_T,H_A,H_L,\Gamma \}$ and the $H_i^{[0,1]}$ are obtained from \eqs{W0final}{W1result}, respectively, corresponding to zeroth and first order in the power expansion. Since we consider $H_i^{[1]}$ at tree level only, we also use the tree-level result for $H_i^{[0]}$ in the denominator for comparison. Note that we keep the full NNLL expressions for the Wilson coefficients $C_i^\incl(q^2)$ in both numerator and denominator (using the numerical expressions from Ref.~\cite{Lee:2006gs}). As already mentioned in \subsec{NLO}, this is consistent because of the split matching and is important for maintaining the correct relative size of the different short distance contributions at $\ord{\lqcd/m_b}$. \fig{Hicorr} shows $\Delta H_T(1,6;m_X^\cut)$, $\Delta H_A(1,3.5;m_X^\cut)$ and $\Delta H_A(3.5,6;m_X^\cut)$, $\Delta H_L(1,6;m_X^\cut)$, and $\Delta \Gamma(1,6;m_X^\cut)$. For $H_L$ and $\Gamma$, the corrections are between $0\%$ and $-10\%$ with central values around $-5\%$ for $m_X^\cut$ between $1.8\GeV$ and $2.0\GeV$. As expected from Ref.~\cite{Lee:2005pw}, the uncertainty in the correction increases for lower $m_X^\cut$. The corrections for $H_T$ are somewhat larger with similar uncertainties. The reason is that in the combination $F_T(k) = F_3(k) - F_4(k)$ entering $H_T$ the corrections from $F_3(k)$ and $F_4(k)$ tend to add up, while in $F_L(k) = F_3(k) + F_4(k)$ entering $H_L$ they tend to cancel.

\begin{figure}[t]
\includegraphics[width=\columnwidth]{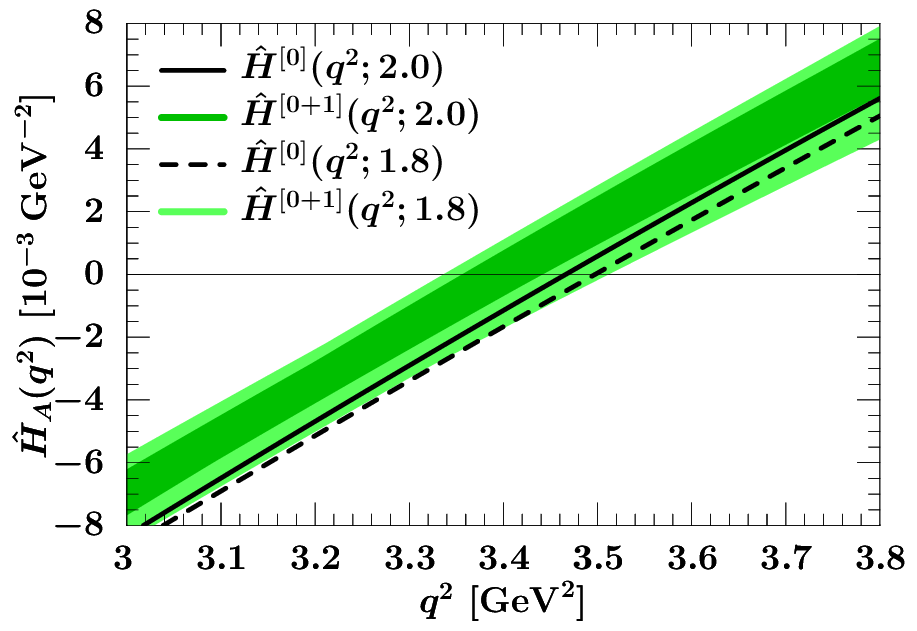}%
\caption{The normalized function $\hat H_A(q^2; m_X^\cut)$ for $m_X^\cut = 2.0\GeV$ (solid and dark green) and $m_X^\cut=1.8\GeV$ (dashed and light green) at lowest order (lines) and including power corrections (bands).}
\label{fig:hatHA}
\end{figure}

Considering $H_A$, we see that the lower bin $H_A(1,3.5)$ receives a significant positive correction, above $+10\%$, while the higher bin $H_A(3.5,6)$ receives only small negative corrections of a few percent. The reason is that the $C_7^\incl(q^2)\,\C_{10}$ term, whose absolute value decreases (it becomes less negative), dominates $H_A$ for very small $q^2$. As $q^2$ increases, these corrections are compensated by a corresponding reduction of the $C_9^\incl(q^2)\,\C_{10}$ term. This also results in a shift of the zero, $q_0^2$, where the forward-backward asymmetry $\df A_{FB}/\df q^2 = (3/4)H_A(q^2)$ vanishes. In \fig{hatHA} we plot the ratio
\begin{equation}
\hat H_A(q^2;m_X^\cut) = \frac{H_A(q^2; m_X^\cut)}{\Gamma(1,6;m_X^\cut)}
\,,\end{equation}
i.e.\ $H_A(q^2)$ normalized to the rate integrated over the low $q^2$ region, as a function of $q^2$ for fixed $m_X^\cut$. The black lines show the leading-order result $\widehat H_A^{[0]}(q^2;m_X^\cut)$ using $m_X^\cut = 2.0\GeV$ (solid) and $m_X^\cut = 1.8\GeV$ (dashed). The green bands show the result obtained by including the subleading shape function corrections in both numerator and denominator, leading to a horizontal shift of about $-0.05\GeV^2$ to $-0.1\GeV^2$ with a similar uncertainty. This is the same size as the perturbative uncertainty usually quoted for $q_0^2$. The size of the horizontal shift in the curve at $q_0^2$ is not different
from that at any other point in this $q^2$ region. This is expected, because in the theoretical description of inclusive decays there is nothing special about the zero beyond the fact that $H_A(q^2)$ happens to vanish there.

\section{Conclusions}
\label{sec:conclusions}

In Ref.~\cite{Lee:2006gs}, it was demonstrated that the three observables $H_T(q^2)$, $H_A(q^2)$, $H_L(q^2)$ measured in the low $q^2$ region provide significantly better sensitivity to the different Wilson coefficients than the rate $\df\Gamma/\df q^2 = H_A(q^2) + H_L(q^2)$ and forward-backward asymmetry $\df A_{FB}/\df q^2 = (3/4)H_A(q^2)$ alone. In the low $q^2$ region, the experimentally required cut on the hadronic invariant mass, $m_X$, makes the measurements sensitive to nonperturbative $b$ quark distribution functions, so-called shape functions. Rather than extrapolating the measurements to compare with theory, one should take the effect of the $m_X$ cut into account on the theory side. In this paper, we computed all three observables, $H_{T,A,L}$, in the low $q^2$ region in the presence of an $m_X$ cut, including the leading and subleading shape function contributions.

We used a split matching procedure to separate the perturbative corrections above and below the scale $m_b$. The perturbative corrections above $m_b$ are taken into account via uniquely defined effective Wilson coefficients $C_i^\incl(q^2)$ and are known at NNLL [$\ord{\alpha_s}$] from the standard calculation of $B\to X_s\ell^+\ell^-$ in the local OPE. Below the scale $\mu_b$, the perturbative corrections at leading order in the power expansion are fully known at NLL [$\ord{\alpha_s}$] and approximately at NNLL [$\ord{\alpha_s^2}$]. The subleading power corrections are included at tree level.

While the effect of the $m_X$ cut at leading order can be taken into account model-independently by combining all constraints on the leading shape function from perturbation theory, together with available data from $B\to X_s\gamma$ and $B\to X_{u,c}\ell\bar\nu$~\cite{Ligeti:2008ac}, much less is known about the subleading shape functions, which represent a currently irreducible hadronic uncertainty. Depending on the observable and the value of the $m_X$ cut, the subleading shape functions induce corrections to the leading-order result of about $-5\%$ to $-10\%$ in the rates and a shift of about $-0.05\GeV^2$ to $-0.1\GeV^2$ in $q_0^2$, with uncertainties of the same size. Hence, they must be accounted for to be able to obtain precise predictions for measurements of $B\to X_s\,\ell^+\ell^-$ in the low $q^2$ region. A detailed numerical analysis of the $m_X$ cut effects and their influence on the uncertainties in the extraction of the Wilson coefficients will be presented in a separate publication~\cite{toappear}.

\acknowledgments

We thank Iain Stewart and Zoltan Ligeti for discussions and comments on the manuscript.
This work was supported in part by the Office of Nuclear Physics of the
U.S.\ Department of Energy (DOE) under the Contract DE-FG02-94ER40818 (F.T.) and in part
by the DOE under the Grant DE-FG02-92ER40701 (K.L.).
K.L. was also supported by the Sherman Fairchild Foundation.


\end{document}